\documentclass{article}

\usepackage{graphicx}
\usepackage{amsmath}

\newtheorem{theorem}{Theorem}
\newtheorem{acknowledgement}[theorem]{Acknowledgement}

\newtheorem{conclusion}[theorem]{Conclusion}

\begin{document}

\title{Kinetic fluctuation in the inhomogeneous plasma}
\author{V.V. Belyi \\
IZMIRAN, Troitsk, Moscow region, 142190, Russia}

\maketitle

\begin{abstract}
Using the Langevin approach and the multiscale technique, a kinetic theory
of the time and space nonlocal fluctuations in the collisional plasma is
constructed. In local equilibrium a generalized version of the Callen-Welton
theorem is derived. It is shown that not only the dissipation but also the
time and space derivatives of the dispersion determine the amplitude and the
width of the spectrum lines of the electrostatic field fluctuations, as well
as the form factor. There appear significant differences with respect to the
non-uniform plasma. In the kinetic regime the form factor is more sensible
to space gradient than the spectral function of the electrostatic field
fluctuations. As a result of the inhomogeneity, these proprieties became
asymmetric with respect to the inversion of the frequency sign. The
differences in amplitude of peaks could become a new tool to diagnose slow
time and space variations in the plasma.


\end{abstract}

\begin{quote}
Fluctuations find an application in diagnostic procedures. Indeed,
plasma parameters such as temperature, mean velocity, density and
their respective profiles can be determined by incoherent
(Thomson) scattering diagnostics \cite{Thomson}, i.e. by the
proper interpretation of data obtained from the scattering of a
given electromagnetic field interacting with the system. The key
point of interpretating them is the knowledge of the intensity of
the dielectric function fluctuations or equally of the electron
form factor $(\delta n_{e}\delta n_{e})_{\omega ,\mathbf{k}}$.
Here $\omega $ and $\mathbf{k}$ are respectively the frequency and
wavevector of the autocorrelations. Due to the Poisson equation
the electron form factor in the spatially homogeneous system is
directly linked to the electrostatic field fluctuations, which
have been the object of active study since the early 1960s
\cite{Thomson}. In the thermodynamic equilibrium, the
electrostatic field fluctuations satisfy the famous
\textbf{Callen-Welton} fluctuation-dissipation theorem \cite{C-W}:
\end{quote}

\begin{equation}
(\delta \mathbf{E}\delta \mathbf{E})_{\omega \mathbf{k}}=\Theta \frac{8\pi
Im\varepsilon (\omega ,\mathbf{k})}{\omega \left| \varepsilon (\omega ,%
\mathbf{k})\right| ^{2}}  \label{c.1}
\end{equation}
linking their intensity to the imaginary part of the dielectric function $%
\varepsilon (\omega ,\mathbf{k})$, and the temperature $\Theta $
in energy units. The spectral function (\ref{c.1}) has peaks,
corresponding to proper plasma frequencies. The matter becomes
more tricky in the non-equilibrium case, when the state of the
plasma is given by Maxwellian distributions characterized by
\emph{different constant} temperatures and velocities per species
$(\Theta _{a},\mathbf{V}_{a};a=e,i)$.
We have indeed shown \cite{B-V}, that, in the \emph{collisional regime}%
\textbf{\ }equations (\ref{c.1}) should be revisited. We stressed
the fact that a \emph{kinetic approach} should be taken.
Introducing fluctuations by the Langevin method, we have
elaborated a ''revisited'' Callen-Welton formula containing,
beside the terms appearing in Eq. (\ref{c.1}), new terms
explicitly displaying dissipative non equilibrium contributions.
\begin{equation}
 (\delta \mathbf{E}\delta \mathbf{E})_{\omega \mathbf{k}%
}=\sum_{a,b=e,i}\frac{8\pi \Theta _{a}}{(\omega -\mathbf{k\cdot V}%
_{a})\left| \varepsilon (\omega ,\mathbf{k})\right| ^{2}}[Im\chi
_{a} + \nu _{ab}(\Theta _{a}-\Theta _{b})\Phi _{1}+\nu _{ab}(\mathbf{%
k\cdot V}_{a}-\mathbf{k\cdot V}_{b})\Phi _{2}],  \label{c.2}
\end{equation}
 where $\chi _{a}$ $(a=e,i)$ is the complex
dielectric susceptibility of the a-th component
 It is important that these new terms contain the interparticle collision
frequency $\nu _{ab}$, the differences in temperatures $(\Theta
_{a}-\Theta
_{b})$ and velocities $(\mathbf{V}_{a}-\mathbf{V}_{b})$, and the functions $%
\Phi _{1}$ and $\Phi _{2}$ of \textbf{real} parts of the dielectric
susceptibilities. It is however not evident that the plasma parameters -
temperature, velocities and densities can be kept \emph{constant}\textbf{\ }%
. Inhomogeneities in space and time of these quantities will
certainly also contribute to the fluctuations. Obviously, to treat
the problem, a kinetic approach is \emph{required} , especially
when the wavelength of the fluctuations is larger than the Debye
wavelength. To derive nonlocal  expressions for the spectral
function of the electrostatic field fluctuation and for the
electron form factor we use the Langevin approach to describe
kinetic fluctuations \cite{Kli, Belyi}. The starting point of our
procedure is the same as in \cite{B-V}. A kinetic equation for the
fluctuation $\delta f_{a}$ of the one-particle distribution
function (DF) with respect to the reference state $f_{a}$ is
considered. In the general case the reference state is a
none-equilibrium DF which varies in space and time both on the
kinetic scale ( mean free path $l_{ei}$ and interparticle
collision time $\nu _{ei}{}^{-1}$) and on the\ larger hydrodynamic
scales. These scales are much larger than the characteristic
fluctuation time $\omega ^{-1}$. In the non-equilibrium case we
can therefore introduce a small parameter $\mu =\nu _{ei}/\omega
$, which allows us to describe fluctuations on the basis of a
multiple space and time scale
analysis. Obviously, the fluctuations vary on both the ''fast'' $(\mathbf{r}%
,t)$ and the ''slow'' $(\mu \mathbf{r},\mu t)$ time and space scales: $%
\delta f_{a}(\mathbf{x,}t)=\delta f_{a}(\mathbf{x},t,\mu t,\mu \mathbf{r})$
and $f_{a}(\mathbf{x,}t)=f_{a}(\mathbf{p},\mu t,\mu \mathbf{r}).$ Here $%
\mathbf{x}$ stands for the phase-space coordinates
$(\mathbf{r,p})$. The Langevin kinetic equation for $\delta f_{a}$
has the form \cite{Kli, B-V}

\begin{equation}
\widehat{L}_{a\mathbf{x}t}(\delta f_{a}(\mathbf{x,}t)-\delta f_{a}^{S}(%
\mathbf{x,}t))=-e_{a}\delta \mathbf{E(r},t\mathbf{)\cdot }\frac{\partial
f_{a}(\mathbf{x,}t)}{\partial \mathbf{p}},  \label{a.3}
\end{equation}

where

$\widehat{\text{ }L}_{a\mathbf{x}t}=\frac{\partial }{\partial t}+\mathbf{v}%
\cdot \frac{\partial }{\partial \mathbf{r}}+\widehat{\Gamma }_{a}(t,\mathbf{r%
});$ $\widehat{\Gamma }_{a}(t,\mathbf{r,p})=e_{a}\mathbf{E\cdot }\frac{%
\partial }{\partial \mathbf{p}}-\delta \widehat{I}_{a};$

$\delta \widehat{I}_{a}$ is the linearized Balescu-Lenard collision operator.

The Langevin source in Eq. (\ref{a.3}) is determined \cite{B-V} by
following equation:

$\widehat{L}_{a\mathbf{x}t}\overline{\delta f_{a}(\mathbf{x,}t)\delta f_{b}(%
\mathbf{x}^{\prime }\mathbf{,}t^{\prime })}^{S}=\delta _{ab}\delta
(t-t^{\prime })\delta (\mathbf{x}-\mathbf{x}^{\prime })f_{a}(\mathbf{x}%
^{\prime }\mathbf{,}t^{\prime }).$

The solution of Eq. (\ref{a.3}) has the form
\begin{equation}
\delta f_{a}(\mathbf{x,}t)=\delta f^{S}(\mathbf{x,}t)-\sum_{b}\int d\mathbf{x%
}^{\prime }\int\limits_{-\infty }^{t}dt^{\prime }G_{ab}(\mathbf{x,}t,\mathbf{%
x}^{\prime }\mathbf{,}t^{\prime })e_{b}\delta \mathbf{E(r}^{\prime
},t^{\prime }\mathbf{)\cdot }\frac{\partial f_{b}(\mathbf{x}^{\prime }%
\mathbf{,}t^{\prime })}{\partial \mathbf{p}^{\prime }},  \label{a.7}
\end{equation}
where the Green function $G_{ab}(\mathbf{x,}t,\mathbf{x}^{\prime }\mathbf{,}%
t^{\prime })$ of the operator $\widehat{L}_{a\mathbf{x}t}$ is determined by $%
\widehat{L}_{a\mathbf{x}t}G_{ab}(\mathbf{x,}t,\mathbf{x}^{\prime }\mathbf{,}%
t^{\prime })=\delta _{ab}\delta (\mathbf{x}-\mathbf{x}^{\prime })\delta
(t-t^{\prime })$

with the causality condition $G_{ab}(\mathbf{x,}t,\mathbf{x}^{\prime }%
\mathbf{,}t^{\prime })=0,$ when $t<t^{\prime }$. Thus, $\overline{\delta
f_{a}(\mathbf{x,}t)\delta f_{b}(\mathbf{x}^{\prime }\mathbf{,}t^{\prime })}%
^{S}$ and $G_{ab}(\mathbf{x,}t,\mathbf{x}^{\prime }\mathbf{,}t^{\prime })$
are connected by the relation:

$\overline{\delta f_{a}(\mathbf{x,}t)\delta f_{b}(\mathbf{x}^{\prime }%
\mathbf{,}t^{\prime })}^{S}=G_{ab}(\mathbf{x,}t,\mathbf{x}^{\prime }\mathbf{,%
}t^{\prime })f_{b}(\mathbf{x}^{\prime }\mathbf{,}t^{\prime }).$

For the stationary and spatially uniform systems, when DF $f_{a}$ and the
operator $\widehat{\Gamma }_{a}$ do not depend on time and space, $G_{ab}(%
\mathbf{x,}t,\mathbf{x}^{\prime }\mathbf{,}t^{\prime })$ can depend only on
its time and space variables through the difference $t-t^{\prime }$ and $%
\mathbf{r}-\mathbf{r}^{\prime }$. In the general case, when the one-particle
DF $f_{a}(\mathbf{p,}\mu \mathbf{r,}\mu t)$ and the operator $\widehat{%
\Gamma }_{a}$ slowly (in comparison with the correlation scales) vary in
time and space, and when non-local effects are considered, the time and
space dependence of $G_{ab}(\mathbf{x,}t,\mathbf{x}^{\prime }\mathbf{,}%
t^{\prime })$ is more subtle.

\begin{equation}
G_{ab}(\mathbf{x,}t,\mathbf{x}^{\prime }\mathbf{,}t^{\prime })=G_{ab}(%
\mathbf{p,p}^{\prime },\mathbf{r-r}^{\prime }\mathbf{,}t\mathbf{-}t^{\prime
},\mu \mathbf{r}^{\prime }\mathbf{,}\mu t^{\prime }).  \label{A.8}
\end{equation}

For the homogeneous case this non-trivial result was obtained for the first
time in \cite{BKW1}. For inhomogeneous systems it has been generalized
recently in \cite{BKW2}.

The relationship (\ref{A.8}) is directly linked with the constitutive
relation between the electric displacement and the electric field:

$D_{i}(\mathbf{r,}t)=\int d\mathbf{r}^{\prime }\int\limits_{-\infty
}^{t}dt^{\prime }\varepsilon _{ij}(\mathbf{r},\mathbf{r}^{\prime
},t,t^{\prime })E_{j}(\mathbf{r}^{\prime },t^{\prime }).$

Previously two kinds of constitutive relations were proposed
phenomenologically for a weakly-inhomogeneous and slowly time-varying medium:

(i) the so-called \textit{symmetrized} constitutive relation \cite{KADOM2}:
\begin{equation}
D_{i}(\mathbf{r,}t)=\int d\mathbf{r}^{\prime }\int \int\limits_{-\infty
}^{t}dt\varepsilon _{ij}(\mathbf{r-r}^{\prime }\mathbf{,}t\mathbf{-}%
t^{\prime };\mu \frac{\mathbf{r+r}^{\prime }}{2}\mathbf{,}\mu \frac{t\mathbf{%
+}t^{\prime }}{2}))E_{j}(\mathbf{r}^{\prime },t^{\prime }).  \label{A.10}
\end{equation}

(ii) the \textit{non-} \textit{symmetrized} constitutive relation \cite
{Silin}:
\begin{equation}
D_{i}(\mathbf{r,}t)=\int d\mathbf{r}^{\prime }\int \int\limits_{-\infty
}^{t}dt\varepsilon _{ij}(\mathbf{r-r}^{\prime }\mathbf{,}t\mathbf{-}%
t^{\prime };\mu \mathbf{r,}\mu t))E_{j}(\mathbf{r}^{\prime },t^{\prime }).
\label{A.11}
\end{equation}
Both phenomenological formulations (i) and (ii) are unsatisfactory. The
correct expression should be

\begin{equation}
D_{i}(\mathbf{r,}t)=\int d\mathbf{r}^{\prime }\int \int\limits_{-\infty
}^{t}dt\varepsilon _{ij}(\mathbf{r-r}^{\prime }\mathbf{,}t\mathbf{-}%
t^{\prime };\mu \mathbf{r}^{\prime }\mathbf{,}\mu t^{\prime }))E_{j}(\mathbf{%
r}^{\prime },t^{\prime }).
\end{equation}

Taking into account the first-order terms with respect to $\mu $ from (\ref
{a.7}) and (\ref{A.8}) we have
\begin{equation*}
\delta f_{a}(\mathbf{x},t)=\delta f_{a}^{S}(\mathbf{x},t)-\sum_{b}\int d%
\mathbf{p}^{\prime }d\mathbf{\rho }\int\limits_{0}^{\infty }d\tau
\end{equation*}
\begin{equation}
\ (1-\mu \tau \frac{\partial }{\partial \mu t}-\mu \mathbf{\rho }\cdot \frac{%
\partial }{\partial \mu \mathbf{r}})e_{b}\delta \mathbf{E}(\mathbf{r}-%
\mathbf{\rho },t-\tau )G_{ab}(\mathbf{\rho },\tau ,\mathbf{p,p}^{\prime
},\mu t,\mu \mathbf{r})\cdot \frac{\partial f_{b}(\mathbf{p}^{\prime },\mu
t,\mu \mathbf{r})}{\partial \mathbf{p}^{\prime }},  \label{A.12}
\end{equation}
$(\mathbf{\rho }=\mathbf{r}-\mathbf{r}^{\prime },\tau =t-t^{\prime }).$

>From the Poisson equation

\begin{equation}  \label{a.19}
\delta \mathbf{E}(\mathbf{r},t)=-\frac \partial {\partial \mathbf{r}%
}\sum\limits_be_b\int \frac 1{\left| \mathbf{r-r}^{\prime }\right| }\delta
f_b(\mathbf{x}^{\prime },t)d\mathbf{x}^{\prime }
\end{equation}

and performing the Fourier-Laplace transformation

$\delta \mathbf{E}(\mathbf{k},\omega )=\int\limits_{0}^{\infty }dt\int d%
\mathbf{r}\delta \mathbf{E}(\mathbf{r},t)\exp (-\Delta t+i\omega t-i\mathbf{k%
}\cdot \mathbf{r}).$

from (\ref{A.12}) we have

\begin{equation*}
\delta \mathbf{E}(\mathbf{k,}\omega ,\mu t,\mu \mathbf{r})=\delta \mathbf{E}%
^{s}(\mathbf{k,}\omega )+\sum\limits_{a}4\pi ie_{a}^{2}\int d\mathbf{p[}(1+i%
\frac{\partial }{\partial \omega }\frac{\partial }{\partial \mu t})
\end{equation*}
\begin{equation*}
\times \frac{\mathbf{k}}{\mathbf{k}^{2}}\widehat{L}_{a\omega \mathbf{k}%
}^{-1}\delta \mathbf{E}(\mathbf{k},\omega ,\mu \mathbf{r},\mu t)\cdot \frac{%
\partial f_{a}(\mathbf{p},\mu \mathbf{r,}\mu t)}{\partial \mathbf{p}}
\end{equation*}
\begin{equation}
-i\frac{\partial }{\partial \mu r_{i}}\delta \mathbf{E}(\mathbf{k},\omega
,\mu \mathbf{r},\mu t)\frac{\partial }{\partial k_{i}}\frac{\mathbf{k}}{%
\mathbf{k}^{2}}\widehat{L}_{a\omega \mathbf{k}}^{-1}\frac{\partial f_{a}(%
\mathbf{p},\mu \mathbf{r,}\mu t)}{\partial \mathbf{p}}].  \label{a.25}
\end{equation}
Here and in the following for simplicity we omit $\mu $, keeping in mind
that derivatives over coordinates and time are taken with respect to the
slowly varying variables.{\large \ }The resolvent in (\ref{a.25}) is
determined by the following relation:

$\int d\mathbf{\rho }\int\limits_{0}^{\infty }d\tau \exp (-\Delta \tau
+i\omega \tau -i\mathbf{k\cdot \rho })G_{ab}(\mathbf{\rho },\tau ,\mathbf{p,p%
}^{\prime }\mathbf{,}\mu t,\mu \mathbf{r})=\widehat{L}_{a\omega \mathbf{k}%
}^{-1}\delta _{ab}\delta (\mathbf{p}-\mathbf{p}^{\prime }).$

The approximation in which Eq. (\ref{a.25}) was derived corresponds to the
geometric optics approximation \cite{Krav}. At first-order and after some
manipulations, one obtains from Eq. (\ref{a.25}) the transport equation in
the geometric optics approximation, which is not considered in the present
article, and the equation for the spectral function of the electrostatic
field fluctuations:
\begin{equation}
Re\varepsilon (\omega ,\mathbf{k})[(\delta \mathbf{E}\delta \mathbf{E}%
)_{\omega ,\mathbf{k}}-\frac{1}{\left| \widetilde{\varepsilon }(\omega ,%
\mathbf{k})\right| ^{2}}(\delta \mathbf{E}\delta \mathbf{E})_{\omega ,%
\mathbf{k}}^{S}]=0,  \label{a.32}
\end{equation}
where we introduced

$\widetilde{\varepsilon }(\omega ,\mathbf{k})=1+\sum\limits_{a}\widetilde{%
\chi }_{a}(\omega ,\mathbf{k});$ $\varepsilon (\omega ,\mathbf{k}%
)=1+\sum\limits_{a}\chi _{a}(\omega ,\mathbf{k})$

\begin{equation}
\widetilde{\chi }_{a}(\omega ,\mathbf{k})=(1+i\frac{\partial }{\partial
\omega }\frac{\partial }{\partial t}-i\frac{\partial }{\partial \mathbf{r}}%
\cdot \frac{\partial }{\partial \mathbf{k}})\chi _{a}(\omega ,\mathbf{k},t,%
\mathbf{r}),  \label{a.29}
\end{equation}
and where

$\chi _{a}(\omega ,\mathbf{k},t,\mathbf{r})=-\frac{4\pi ie_{a}^{2}}{k^{2}}%
\int d\mathbf{p}\widehat{L}_{a\omega \mathbf{k}}^{-1}\mathbf{k}\cdot \frac{%
\partial }{\partial \mathbf{p}}f_{a}(\mathbf{p,}t,\mathbf{r})$

is the susceptibility for a collisional plasma. In the same approximation
the spectral function of the Langevin source $(\delta \mathbf{E}\delta
\mathbf{E})_{\omega ,\mathbf{k}}^{S}$ takes the form
\begin{equation}
(\delta \mathbf{E}\delta \mathbf{E})_{\omega ,\mathbf{k}}^{S}=32\pi
^{2}\sum\limits_{a}e_{a}^{2}Re\int d\mathbf{p}(1+i\frac{\partial }{\partial
\omega }\frac{\partial }{\partial t}-i\frac{\partial }{\partial \mathbf{k}}%
\cdot \frac{\partial }{\partial \mathbf{r}})\frac{1}{k^{2}}\widehat{L}%
_{a\omega \mathbf{k}}^{-1}f_{a}(\mathbf{p},\mathbf{r},t).  \label{a.32a}
\end{equation}

If $Re\mathbf{\varepsilon (}\omega \mathbf{,k)}\neq 0$, it follows from Eqs.
(\ref{a.32}) and (\ref{a.32a})that the spectral function of the
nonequilibrium electrostatic field fluctuations is determined by the
expression:

\begin{equation}
\lbrack (\delta \mathbf{E}\delta \mathbf{E})_{\omega ,\mathbf{k}}=\frac{%
32\pi ^{2}\sum\limits_{a}e_{a}^{2}Re\int d\mathbf{p}(1+i\frac{\partial }{%
\partial \omega }\frac{\partial }{\partial t}-i\frac{\partial }{\partial
\mathbf{k}}\cdot \frac{\partial }{\partial \mathbf{r}})\frac{1}{k^{2}}%
\widehat{L}_{a\omega \mathbf{k}}^{-1}f_{a}(\mathbf{p},\mathbf{r},t)}{\left|
\widetilde{\varepsilon }(\omega ,\mathbf{k})\right| ^{2}}.  \label{a.29a}
\end{equation}
The effective dielectric function $\widetilde{\varepsilon }(\omega ,\mathbf{k%
})$ in the denominator of Eq. (\ref{a.29a}) determines the spectral
properties of the electrostatic field fluctuations and its imaginary part

\begin{equation}
Im\widetilde{\varepsilon }(\omega ,\mathbf{k})=Im\varepsilon (\omega ,%
\mathbf{k})+\frac{\partial }{\partial \omega }\frac{\partial }{\partial t}%
Re\varepsilon (\omega ,\mathbf{k},t,\mathbf{r})-\frac{\partial }{\partial
\mathbf{k}}\cdot \frac{\partial }{\partial \mathbf{r}}Re\varepsilon (\omega ,%
\mathbf{k},t,\mathbf{r}),  \label{a.29b}
\end{equation}
determines the width of the spectral lines near the resonance. Note that
when expanding the Green function in Eq. (\ref{A.12}) in terms of the small
parameter $\mu $, there appear additional terms at first order. It is
important to note that the\emph{\ imaginary} part of the dielectric
susceptibility is now replaced by the \emph{real} part, which is greater
than \emph{imaginary}\textit{\ }part by the factor $\mu ^{-1}$. Therefore,
the second and third terms in Eq. (\ref{a.29b}) in the kinetic regime have
an effect comparable to that of the first term. At second order in the
expansion in $\mu $ the corrections appear only in the \emph{imaginary }part
of the susceptibility, and they can reasonably be neglected. It is therefore
sufficient to retain the first order corrections to solve the problem.

For the local equilibrium case where the reference state $f_{a}$ is
Maxwellian, we have the identity: $\int d\mathbf{p}(1+i\frac{\partial }{%
\partial \omega }\frac{\partial }{\partial t}-i\frac{\partial }{\partial
\mathbf{k}}\mathbf{\cdot }\frac{\partial }{\partial \mathbf{r}})\frac{1}{%
k^{2}}\widehat{L}_{a\omega \mathbf{k}}^{-1}f_{a}(\mathbf{p},t,\mathbf{r})=%
\frac{i}{\omega _{a}}\int d\mathbf{p}f_{a}(\mathbf{p},t,\mathbf{r})-\frac{%
i\Theta _{a}}{\omega _{a}4\pi e_{a}^{2}}\widetilde{\chi }_{a}(\omega ,%
\mathbf{k})$ $(\omega _{a}=\omega -\mathbf{kV}_{a}$) and Eq.(\ref{a.29a})
takes the form
\begin{equation}
(\delta \mathbf{E}\delta \mathbf{E})_{\omega ,\mathbf{k}}=\sum_{a}\frac{8\pi
\ \Theta _{a}}{\omega _{a}\left| \widetilde{\varepsilon }\mathbf{(}\omega ,%
\mathbf{k})\right| ^{2}\mathbf{\ }}Im\widetilde{\chi }_{a}(\omega ,\mathbf{k}%
).  \label{a.38}
\end{equation}

In this case the small parameter $\mu $ is determined on the slower
hydrodynamic scale. For the case of equal temperatures and $\mathbf{V}_{a}=0$
one obtains a generalized expression for the Callen-Welton formula:
\begin{equation}
(\delta \mathbf{E}\delta \mathbf{E})_{\omega ,\mathbf{k}}=\frac{8\pi \
\Theta Im\widetilde{\varepsilon }\mathbf{(}\omega ,\mathbf{k})}{\omega
\left| \widetilde{\varepsilon }\mathbf{(}\omega ,\mathbf{k})\right| ^{2}%
\mathbf{\ }}.  \label{a.39}
\end{equation}

To calculate explicitly $(\delta \mathbf{E}\delta \mathbf{E})_{\omega ,%
\mathbf{k}}$ we will restrict our analysis to the vicinity of the resonance,
\textit{i.e. }$\omega =\pm \omega _{0}$, where $Re\varepsilon \mathbf{(}%
\omega _{0},\mathbf{k})=0$. We can develop $\widetilde{\varepsilon }(\omega ,%
\mathbf{k})=(\omega -\omega _{0}sgn\omega )\frac{\partial Re\varepsilon }{%
\partial \omega }\lfloor _{\omega =\omega _{0}sgn\omega }$ $+i[Im\varepsilon
+(\frac{\partial ^{2}}{\partial \omega \partial t}-\frac{\partial }{\partial
\mathbf{k}}\cdot \frac{\partial }{\partial \mathbf{r}})Re\varepsilon
]\lfloor _{\omega =\omega _{0}sgn\omega }.$ Thus $(\delta \mathbf{E}\delta
\mathbf{E})_{\omega ,\mathbf{k}}=\frac{\widetilde{\gamma }}{(\omega -\omega
_{0}sgn\omega )^{2}+\widetilde{\gamma }^{2}}\frac{8\pi T}{\omega \partial
Re\varepsilon /\partial \omega }\lfloor _{\omega =\omega _{0}},$ where
\begin{equation}
\widetilde{\gamma }=[Im\varepsilon +\frac{\partial ^{2}}{\partial \omega
\partial t}Re\varepsilon -\frac{\partial }{\partial \mathbf{k}}\cdot \frac{%
\partial }{\partial \mathbf{r}}Re\varepsilon ]/\frac{\partial Re\varepsilon
}{\partial \omega }\lfloor _{\omega =\omega _{0}sgn\omega }  \label{a.42}
\end{equation}
is the effective damping decrement. For the case where the system
parameters are homogeneous in space but vary in time, the
correction is still symmetric with respect to the change of sign
of $\omega $, but the intensities and broadening are different,
and the intensity integrated over the frequencies remains the same
as in the stationary case. However, when the plasma parameters are
space dependent this symmetry is lost. The spectral asymmetry is
related to the appearance of space anisotropy in inhomogeneous systems. The\textit{\ }%
\emph{real} part of the susceptibility $Re\varepsilon $ is an even function
of $\omega $. This property implies that the contribution of the third term
to the expression of the damping decrement (\ref{a.42}) is an odd function
of $\omega $. Moreover this term gives rise to an anisotropy in $k$ space.

Let us estimate this correction for the plasma mode $(\omega _{0}=\omega
_{L})$ $Re\varepsilon =1-\frac{\omega _{L}^{2}}{\omega ^{2}}(1+3\frac{%
k^{2}\Theta }{m\omega ^{2}}),$ $Im\varepsilon =\frac{\omega _{L}^{2}}{\omega
^{2}}\frac{\nu _{ei}}{\omega },$ $\omega _{L}^{2}=\frac{4\pi ne^{2}}{m}=%
\frac{\Theta k_{D}^{2}}{m}$ and
\begin{equation}
\widetilde{\gamma }=[\nu _{ei}+\frac{2}{n}\frac{\partial n}{\partial t}+6%
\frac{\omega _{L}}{nk_{D}^{2}}\mathbf{k\cdot }\frac{\partial n}{\partial
\mathbf{r}}sgn\omega ]/2.  \label{a.44}
\end{equation}

For the spatially homogeneous case there is no difference between the
spectral properties of the longitudinal electric field and of the electron
density. They are connected by the Poisson equation. This statement is no
longer valid when considering an inhomogeneous plasma. Indeed the
longitudinal electric field is linked to the particle density by the
nonlocal Poisson relation (\ref{a.19}). In the latter case, an analysis
similar to that made above can also be performed for the particle density.
>From Eq. (\ref{a.7}) there follows
\begin{equation*}
\delta n_{a}(\mathbf{k,}\omega ,\mathbf{r,}t)=\delta n_{a}^{S}(\mathbf{k,}%
\omega ,\mathbf{r,}t)+\sum\limits_{b}\frac{4\pi i\mathbf{k}e_{b}e_{a}}{k^{2}}%
\int d\mathbf{p[}(1+i\frac{\partial }{\partial \omega }\frac{\partial }{%
\partial t})\widehat{L}_{a\omega \mathbf{k}}^{-1}\delta n_{b}(\mathbf{k}%
,\omega ,\mathbf{r},t)\cdot \frac{\partial f_{a}(\mathbf{p,r},t)}{\partial
\mathbf{p}}
\end{equation*}

\begin{equation}
-i\frac{\partial }{\partial r_{i}}\delta n_{b}(\mathbf{k},\omega ,\mathbf{r}%
,t)\frac{\partial }{\partial k_{i}}\widehat{L}_{a\omega \mathbf{k}}^{-1}%
\frac{\partial f_{a}(\mathbf{p,r},t)}{\partial \mathbf{p}}].  \label{a.47}
\end{equation}
At the first order approximation and after some manipulations, one obtains
the following expression for the electron form factor for a two-component ($%
a=e,i$) plasma:
\begin{equation*}
(\delta n_{e}\delta n_{e})_{\omega ,\mathbf{k}}=\frac{2n_{e}k^{2}}{\omega
_{e}k_{D}^{2}}\frac{\left| 1+\widetilde{\widetilde{\chi _{i}}}(\omega ,%
\mathbf{k)}\right| ^{2}}{\left| \widetilde{\widetilde{\varepsilon }}(\omega
\mathbf{,k)}\right| ^{2}}Im\widetilde{\widetilde{\chi _{e}}}(\omega ,\mathbf{%
k})
\end{equation*}

\begin{equation}
+\left| \frac{\widetilde{\widetilde{\chi _{e}}}\mathbf{(}\omega ,\mathbf{k)}%
}{\widetilde{\widetilde{\varepsilon }}(\omega \mathbf{,k)}}\right| ^{2}\frac{%
\Theta _{i}}{\Theta _{e}}\frac{2n_{e}k^{2}}{\omega _{i}k_{D}^{2}}Im%
\widetilde{\widetilde{\chi _{i}}}(\omega ,\mathbf{k}),
\end{equation}
where \ we used for local equilibrium the following expression for the
''source'' $(\delta n_{a}\delta n_{b})_{\omega ,\mathbf{k}}^{S}=\delta _{ab}%
\frac{\Theta _{a}}{\omega _{a}}\frac{k^{2}}{2\pi e_{a}^{2}}Im\widetilde{%
\widetilde{\chi _{a}}}(\omega ,\mathbf{k})$,

and $\widetilde{\widetilde{\varepsilon }}(\omega \mathbf{,k)}%
=1+\sum\limits_{a}\widetilde{\widetilde{\chi _{a}}}(\omega ,\mathbf{k});%
\widetilde{\widetilde{\chi _{a}}}(\omega ,\mathbf{k})=(1+i\frac{\partial }{%
\partial \omega }\frac{\partial }{\partial t}-i\frac{1}{k^{2}}\frac{\partial
}{\partial r_{i}}k_{j}\frac{\partial }{\partial k_{i}}k_{j})\chi _{a}(\omega
,\mathbf{k},t,\mathbf{r}).$ As above we can expand $\widetilde{\widetilde{%
\varepsilon }}(\omega \mathbf{,k)}$ near the plasma resonance $\omega
=\omega _{L}$. Thus, for the electron line,

$(\delta n_{e}\delta n_{e})_{\omega ,\mathbf{k}}=\frac{\widetilde{\widetilde{%
\gamma }}}{(\omega -sign\omega _{L})^{2}+(\widetilde{\widetilde{\gamma }}%
)^{2}}\frac{2n_{e}k^{2}}{\omega k_{D}^{2}\partial Re\varepsilon /\partial
\omega }\lfloor _{\omega =\omega _{L}}$,

where
\begin{equation}
\widetilde{\widetilde{\gamma }}=[Im\varepsilon +\frac{\partial
^{2}Re\varepsilon }{\partial t\partial \omega }-\frac{1}{k^{2}}\frac{%
\partial }{\partial r_{i}}k_{j}\frac{\partial }{\partial k_{i}}%
k_{j}Re\varepsilon ]/\frac{\partial Re\varepsilon }{\partial \omega }\lfloor
_{\omega =\omega _{L}sgn\omega }  \label{a.66}
\end{equation}
is the effective damping decrement for the electron form factor. At this
stage of calculation, let us note that the damping decrements for the
electrostatic field fluctuations [Eq. (\ref{a.42})] and for the electron
density fluctuations [Eq. (\ref{a.66})] are not the same. The origin of this
difference is that the Green function for electrostatic field fluctuation
and density particle fluctuations are not the same. This property holds only
in the inhomogeneous situation. An estimation for the plasma mode is then:
\begin{equation}
\widetilde{\widetilde{\gamma }}=[\nu _{ei}+\frac{2}{n}\frac{\partial n}{%
\partial t}+\frac{\omega _{L}}{nk^{2}}\mathbf{k\cdot }\frac{\partial n}{%
\partial \mathbf{r}}(1+\frac{6k^{2}}{k_{D}^{2}})sgn\omega ]/2.  \label{a.68}
\end{equation}
>From this equation we see that the inhomogeneous correction in Eq.(\ref{a.68}%
) is greater than the one in Eq. (\ref{a.44}) by the factor $%
1+k_{D}^{2}/6k^{2}$. For the same inhomogeneity; i.e., the same gradient of
the density, we plot the form factor $(\delta n_{e}\delta n_{e})_{\omega ,%
\mathbf{k}}$ together with the $(\delta \mathbf{E}\delta \mathbf{E})_{\omega
,\mathbf{k}}$ as functions of frequency (Fig. 1). This figure shows that the
asymmetry of the spectral lines is present both for $(\delta n_{e}\delta
n_{e})_{\omega ,\mathbf{k}}$ and $(\delta \mathbf{E}\delta \mathbf{E}%
)_{\omega ,\mathbf{k}}$. However, this effect is more pronounced in $(\delta
n_{e}\delta n_{e})_{\omega ,\mathbf{k}}$ than in $(\delta \mathbf{E}\delta
\mathbf{E})_{\omega ,\mathbf{k}}$.

\begin{conclusion}
We have shown that the amplitude and the width of the spectral lines of the
electrostatic field fluctuations and form factor are affected by new
non-local dispersive terms. They are not related to Joule dissipation and
appear because of an additional phase shift between the vectors of induction
and electric field. This phase shift results from the finite time needed to
set the polarization in the plasma with dispersion. Such a phase shift in
the plasma with space dispersion appears due to the medium inhomogeneity.\
These results are important for the understanding and the classification of
the various phenomena that may be observed in applications; in particular,
the asymmetry of lines can be used as a diagnostic tool to measure local
gradients in the plasma.
\end{conclusion}

\begin{acknowledgement}
I acknowledge support from Russian Foundation for Basic Research
(grant 03-02-16345).
\end{acknowledgement}

\begin{figure}
\begin{center}
  \includegraphics{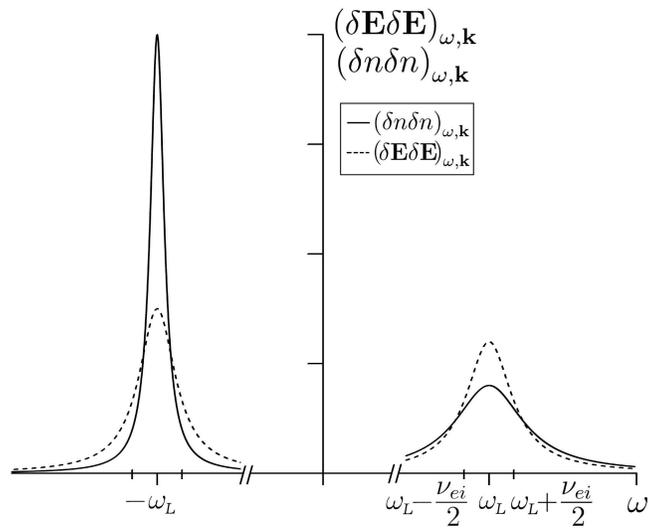}\\
  \caption{The electron form factor $(\delta n_{e}\delta n_{e})_{\omega ,%
\mathbf{k}}$ ( solid line) and the spectral function of electrostatic field
fluctuations $(\delta \mathbf{E}\delta \mathbf{E})_{\omega ,\mathbf{k}}$
(dashed line) as a function of frequency. $\mathbf{k\cdot }\frac{\partial n}{%
\partial \mathbf{r}}=\frac{\nu _{ei}nk_{D}^{2}}{54\omega _{L}};$ $\frac{k_{D}%
}{k}=6$}
  \end{center}
\end{figure}

\end{document}